 \documentstyle[11pt,epsf]{article}

 \renewcommand{\d}{{\mathrm d}} 

\begin{document}
 
 \baselineskip 18pt

 \begin{titlepage}

 \begin{flushright}
   SHEP 97-09\\
   UM-TH-97-12 \\
   hep-ph/9705466\\
   29 May 1997
 \end{flushright}
 
 \vspace{.4in}

 \begin{center}
   {\large{\bf Running coupling and Borel singularities at small $x$}} \\
   \bigskip
   K.D. Anderson$^1$, D.A. Ross$^1$, M.G. Sotiropoulos$^2$. \\
   \bigskip
   {\it $^1$Physics Department, University of Southampton \\
     Southampton, SO17 1BJ, U.K. \\
     $^2$ Randall Laboratory, University of Michigan \\
     Ann Arbor, MI 48109, U.S.A. }
 \end{center}
 
 \vspace{.5in}

 \begin{abstract}
   
   Starting from the dipole representation of small-$x$ evolution we
   implement the running of the coupling in a self-consistent way.
   This results in an evolution equation for the dipole density in
   Borel $(b)$ space.  We show that the Borel image of the dipole
   density is analytic in the neighbourhood of $b=0$ and that it is
   equal to the BFKL solution at $b=0$.  We study the Borel
   singularity structure of the dipole cascade emanating from a
   virtual photon at small $x$ and find a branch cut on the positive
   $b$-semiaxis starting at $b=1/ \beta_0$.  This indicates the
   presence of $1/Q^2$ power corrections to the small-$x$ structure
   functions.  Finally we present numerical results in the context of
   D.I.S.
   
   \bigskip

 \noindent
 PACS numbers: 11.55.J, 12.40.M \hfill

 \end{abstract}
 
 \setcounter{page}{0}

 \end{titlepage}

 \newpage

 \section{Introduction}
 
 The study of QCD in the high energy limit has a history spanning at
 least twenty years and is still actively pursued.  Within
 perturbation theory (pQCD) an understanding has emerged concerning
 hadronic processes in the semihard regime, such as near-forward
 elastic and diffractive scattering, small-$x$ deeply inelastic
 scattering and rapidity gap events.  The evolution of the partonic
 cascade stemming from the initial state hadrons is known to be
 described in the leading logarithmic approximation LLA(x) by the BFKL
 equation either in its original form \cite{BFKL} or in its improved
 reformulations \cite{kTfac}.  However, the picture is yet incomplete
 because of two major problems encountered in this regime.  The first
 has to do with unitarity \cite{Muelunit}, which is violated by the
 LLA(x) result.  The second concerns the infrared sensitivity of the
 LLA(x) result, which is infrared finite but does receive
 contributions from low momentum regions, where physical observables
 are sensitive to large non-perturbative corrections.  It is the
 second problem that we shall consider in this paper.
 
 Unlike the more conventional high momentum transfer ($Q$) processes
 with final state particles in the central rapidity region, the
 semihard kinematic regime is characterised by three ordered scales, $
 \sqrt{s} \gg Q \gg \Lambda_{\mathrm QCD}$.  Note that the use of pQCD is
 justified by the second of the previous inequalities, in the absence
 of which, Regge phenomenology of soft processes is the only available
 approach so far.  The pQCD factorisation theorem, applicable for
 high-$Q$ processes, states that inclusive observables can be
 expressed as a product of parton distribution functions and a hard
 scattering cross section (matrix elements and coefficient function in
 OPE parlance).  In the semihard regime this theorem needs amendment.
 Here, the product is not only a convolution in longitudinal momentum
 fractions of the partons but also in their transverse momentum.  This
 is known as $k_T$-factorisation \cite{kTfac} and in impact parameter
 space it translates to a convolution in impact parameter of the
 radiated secondary partons and an off-shell hard amplitude or cross
 section.  This leads us to the first of the two theoretical inputs of
 the present study, namely the use of the impact parameter or dipole
 emission formulation of semihard processes, as developed by Mueller
 \cite{Mueller} and independently by Nikolaev and Zakharov
 \cite{NikZak2}.  The merit of this formalism, apart from the
 simplicity of the final results, is that it organises the process in
 terms of sequential soft gluon emissions in the $s$-channel, which in
 the Coulomb gauge admit a clear physical interpretation.  They are to
 be understood as initial state radiation occurring before the hard
 scattering.  Final state interactions are not included in the
 evolution equation because they only lead to a unitarity
 rearrangement that does not affect inclusive observables.
 
 To study the infrared sensitivity of a semihard process we are bound
 to consider a subset of next-to-leading logarithmic corrections,
 specifically the ones that have to do with the running coupling.  In
 the BFKL approach, the evolution kernel, although infrared finite, is
 sensitive to the running coupling because of diffusion towards the
 infrared of the transverse momentum along the exchanged gluon ladder.
 With specific assumptions about the infrared dynamics (i.e. long
 wavelength gluon propagation) and the scale of the running coupling,
 the effect on the Regge trajectories has been analyzed in
 refs.~\cite{LHR}.  In the present study we take the point of view
 that in the dipole formalism, just like in timelike parton cascades,
 the scale of the running coupling is set by the virtuality of the
 emitted gluon.  The problem of translating this into impact parameter
 space is resolved through the use of the Borel transformation.  We
 require that the dipole evolution kernel has the same Borel
 singularity structure as the one obtained in momentum space.  This
 condition is sufficient to determine the scale of the coupling as a
 function of the impact parameters of the emitted gluon.  Therefore,
 the second theoretical input in our approach is renormalon analysis
 \cite{renormalons}.  Once the dipole kernel with running coupling is
 constructed, we study the power corrections that are generated when
 this kernel is used to evolve the wave function of an initial
 hadronic state such as a virtual photon in small-$x$ D.I.S.
 
 The structure of this paper is as follows.  In section 2 we highlight
 the main concepts and results of the dipole formalism.  Here we also
 review the derivation of the dipole evolution equation with running
 coupling.  In section 3 we discuss the Borel singularity structure of
 the dipole density.  
 We find a series of infrared renormalon singularities at
 $b \beta_0 = n$, $n=1,2,3,...$ as well as a leading singularity at
 $b \beta_0 =\gamma$, with $0<\gamma<1$ the anomalous dimension of the
 solution.  We show that the Borel image of the dipole kernel is
 analytic in the neighbourhood of $b=0$ and compute its action as a
 power series expansion in $b$.  In the limit $b \rightarrow 0$ the
 BFKL result is recovered.  Iterations of the kernel turn the
 renormalon poles into branch cuts but do not shift their positions.
 In section 4 we discuss small-$x$ D.I.S. in the dipole formalism using
 the results derived previously.  The main conclusion is that the
 leading Borel singularity is at $b=1/\beta_0$, which implies $1/Q^2$
 power corrections consistent with the Wilson OPE expectation.  The
 dipole evolution equation for fixed coupling can be solved by virtue
 of its scale invariance and leads to the same spectrum as that of the
 BFKL equation.  For the Borel image this scale invariance is
 necessarily lost and the resulting evolution equation can only be
 solved in numerical approximation by successive iterations.  In
 section 5 we present results for the first few iterations of the
 Borel transformed kernel in deeply inelastic scattering, and demonstrate that
 the singularity conforms with that discussed in section 4.  We
 summarise in the final section.

 \section{The dipole kernel with running coupling}
 
 We consider small-$x$ D.I.S. as a specific example of a semihard
 process and describe it in the rest frame of the nucleon target of
 mass $ m_N$.  Let $\psi^{(0)}(z, {\mathbf r})$ denote the wave
 function for $\gamma^\star$ to fluctuate into a $q$-$\overline{q}$
 dipole of transverse size $r$ and with the quark carrying
 longitudinal momentum fraction $z$.  Then, the virtual
 photoabsorption cross section $\sigma(x, Q^2)$ for given polarisation
 of $\gamma^\star$ $(T,L)$, can be written as \cite{NikZak1, NikZak3}
 \begin{equation} \sigma_{T,L}(x, Q^2) = \int_0^1 \d z \int \d^2 {\mathbf r}
 |\psi^{(0)}_{T,L}(z, {\mathbf r})|^2 \sigma_{d N}(Y, {\mathbf r}) \,
 ,
 \label{impfac}
 \end{equation} where $\sigma_{d N}$ denotes the dipole-nucleon total cross
 section and $Y= \ln(z/x)$ is the large rapidity parameter to be
 resummed within pQCD.  The physical interpretation of
 eq.~(\ref{impfac}) is that the transition $\gamma^\star \rightarrow q
 \overline{q}$, with light cone time scale $\tau = {\mathcal O}(1/x
 m_N)$, occurs much earlier than subsequent interactions with the
 nucleon target.  The lowest order in $\alpha_s$ dipole cross section
 $\sigma^{(0)}_{d N}$ is \begin{equation} \sigma^{(0)}_{d N}({\mathbf r}) =
 \frac{1}{N_c} \int \frac{\d^2 {\mathbf l}}{({\mathbf l}^2)^2}
 \alpha_s^2 V({\mathbf l}) Re (1-e^{-i {\mathbf l} {\mathbf r}}) \, ,
 \label{losigma}
 \end{equation} with $V({\mathbf l})$ proportional to the absorptive part of the
 gluon-gluon-nucleon-nucleon vertex function.
 
 The LLA(x) radiative corrections are generated by emission of soft
 gluons by the initial $q$-$\overline{q}$ dipole with longitudinal
 momentum fractions strongly ordered as \begin{equation} z \gg z_1 \gg z_2 \gg
 ...\gg z_n \, .
 \label{ordered}
 \end{equation} The light cone time scale for the $k$-th gluon emission is
 $\tau_k = {\mathcal O}(z_k Q/ {\mathbf k}_k^2)$, therefore it is
 emitted independently of the previous ones $(\tau_k \gg \tau_{k+1})$.
 The $\gamma^\star$ wave function $\psi^{(n)}(z, z_i,{\mathbf r},
 {\mathbf r}_i)$ for the emission of $n$ soft gluons with longitudinal
 momentum fractions $z_i$ and at transverse positions ${\mathbf r}_i$
 can be calculated order by order in pQCD and the inclusive
 probability distribution for $\gamma^\star$ to fluctuate into
 $q$-$\overline{q}$ plus soft radiation, $\Phi(z, r)$, is obtained
 by the square of $\psi^{(n)}$ integrated over the phase space of the
 emitted gluons and summed and averaged over their polarisations
 $\{\lambda_i\}$ and colours $\{ a_i\}$.  
 \begin{eqnarray} 
 \Phi^{(n)}(z, r) &=&
 \sum_{ \lambda_i, a_i} \, \prod_{i=1}^n \frac{\d^2 {\mathbf r}_i}{2
   \pi} \, \int^{z_{i-1}} \frac{\d z_i}{2 z_i} \, |\psi_{
   \{\lambda_i\}}^{(n) \{a_i\}}(z, z_1, ... z_n ,
 {\mathbf r}, {\mathbf r}_1, ... {\mathbf r}_n) |^2 \, ,   \\
 \Phi^{(0)}(z, r) &=& |\psi^{(0)}(z, {\mathbf r})|^2 \, .  \nonumber
 \label{Phidef}
 \end{eqnarray} 
 Here and until section 4 we suppress the $\gamma^\star$
 polarisation indices $(T, L)$ for brevity. 
Indeed it is the projection of these polarisation states for the virtual
 photon that is responsible for the fact that
 $ \Phi^{(0)}(z, r)$ is
 taken to be a function of $r$ only.

In the Coulomb gauge the soft radiation can be viewed as a cascade of
 colour dipoles emanating from the initial $q$-$\overline{q}$ dipole.
 It is useful to introduce the dipole density $n(Y, r, \rho)$
 \cite{Mueller, MuellPatel}, for emission from the initial
 $q$-$\overline{q}$ dipole of size $r$ of a dipole of size $\rho$ and
 with the smallest longitudinal momentum fraction in the emitted
 dipole bounded from below by $e^{-Y}$.  The dipole density is
 independent of the projectile bound state properties.  When
 convoluted with the $\gamma^\star$ wave function squared $\Phi^{(0)}$
 it generates the number density $N(Y, \rho)$ of dipoles of size
 $\rho$ and rapidity larger than Y as 
 \begin{equation} N(Y, \rho) = \int_0^1 \d z
 \, \int \d^2 {\mathbf r} \, \Phi^{(0)}(z, r) \, n(Y, r, \rho) \, .
 \label{numberdensity}
 \end{equation} 
 The whole dipole cascade can be constructed from the repeated
 action of a kernel ${\mathcal K}$ on the initial density $n_0(Y,
 r,\rho)$ through the dipole evolution equation 
 \begin{equation} n(Y, r, \rho) =
 n_0(r, \rho) + \int_0^Y \d y \int_0^\infty \d r^\prime {\mathcal
   K}(r, r^\prime) \, n(y, r^\prime, \rho) \, .
 \label{dipev}
 \end{equation} 
 with driving term \begin{equation} n_0( r, \rho) = n(Y=0, r, \rho) =
 r \delta(r - \rho) \, .
 \label{ninit}
 \end{equation}
 
 The evolution kernel ${\mathcal K}$ can be calculated from the one
 soft gluon emission probability $\Phi^{(1)}(z, r)$.  The result is
 \begin{equation} \Phi^{(1)}(z, r) = \Phi^{(0)}(z, r) \, 
 \int^z \, \frac{\d z_1}{z_1} \int \, \d^2
 {\mathbf r}_1 \, \frac{\alpha_s C_F}{ \pi^2} \, \frac{ r^2}{r_1^2 \,
   \hat{r}_1^2} \, , 
 \label{Phi1}
 \end{equation} 
 where hatted vectors will always denote the distance from the
 second parent emitter, ${\mathbf \hat{r}}_1 \, := {\mathbf r}_1 -
 {\mathbf r}$.  
 The full kernel ${\mathcal K}$ is 
 \begin{equation} {\mathcal
   K}(r, r^\prime) = -\frac{\alpha_s C_F}{\pi^2} \, \delta(r
 -r^\prime) \int \d^2 {\mathbf r}^{\prime \prime} \frac{r^2} {
   r^{\prime \prime 2} \, \hat{r}^{\prime \prime 2}} + 2\frac{\alpha_s
   C_F}{\pi^2}\, \int_0^\infty \d \hat{r}^\prime J(r, r^\prime,
 \hat{r}^\prime) \frac{r^2}{ r^{\prime 2} \, \hat{r}^{\prime 2}} \, .
 \label{kernel}
 \end{equation} 
 The first term of ${\mathcal K}$ comes from the one loop virtual
 corrections to the initial $q$-$\overline{q}$ dipole of size $r$.
 The $\delta$-function denotes the absence of gluon radiated into the
 final state.  The second term corresponds to real emission and can be
 read off from eq.~(\ref{Phi1}) taking into account eq.~(\ref{dipev}).
 A factor of 2 is included here to account for the fact that after one
 dipole splits into two, each one of the offsprings can act as parent
 for subsequent emissions.  The change of integration variables 
 \begin{equation}
 \d^2 {\mathbf r}\: ^\prime = J(r, r^\prime, \hat{r}^\prime) \, \d
 r^\prime \, \d \hat{r}^\prime
 \label{chvars}
 \end{equation} 
 generates the (triangle) Jacobian \cite{Mueller} 
 \begin{equation} J(r, r^\prime,
 \hat{r}^\prime) = 2 \pi \, r^\prime \, \hat{r}^\prime \,
 \int_0^\infty \d \kappa \, \kappa \, J_0(\kappa r) \, J_0(\kappa
 r^\prime) \, J_0( \kappa \hat{r}^\prime) \, .
 \label{jacobian}
 \end{equation}
 
 Unlike the BFKL kernel, both virtual and real parts of the dipole
 kernel ${\mathcal K}$ are separately IR finite in the limit
 $r^\prime, \hat{r}^\prime \rightarrow \infty$.  For fixed $\alpha_s$
 and in the limit $N_c \rightarrow \infty$, for which $2 C_F
 \rightarrow N_c$, ${\mathcal K}$ is conformally invariant and has the
 same spectrum as the BFKL kernel, i.e.  
 \begin{equation} \int_0^\infty \d
 r^\prime {\mathcal K}(r, r^\prime) (r^{\prime 2})^ \gamma =
 \frac{\alpha_s N_c}{\pi} \chi(\gamma) (r^2)^\gamma \, ,
 \label{eigeneq}
 \end{equation} where \begin{equation} \chi(\gamma) = 2 \Psi(1) - \Psi(\gamma) -
 \Psi(1-\gamma) \, ,
 \label{chifn}
 \end{equation} 
 is the BFKL spectral function.  Indeed, the two approaches lead
 to the same phenomenological results for inclusive observables.  The
 equivalence between the dipole and BFKL formalisms at the level of
 light cone perturbation theory diagrams has been shown explicitly in
 ref. \cite{ChenMuel}.
 
 After the introduction of the dipole density $n$, the dipole cross
 section $\sigma_{dN}$ of eq.~(\ref{impfac}) takes the form 
 \begin{equation}
 \sigma_{d N}(Y,  r) = \int \frac{\d^2 {\mathbf \rho}}{ 2 \pi
   \rho^2} \, n(Y, r, \rho) \, \sigma_{dN}^{(0)}(\rho) \, .
 \label{sigman}
 \end{equation} 
 Eqs.~(\ref{impfac}, \ref{sigman}) provide the factorised form of
 small-$x$ D.I.S. in impact parameter space.  We note that 
 dependence on target mass scale $m_N$ enters through $\sigma^{(0)}_{d N}$ 
 and that the universal part of the photoabsorption cross section 
 $\sigma(x, Q^2)$ is indeed the dipole density $n$,
 which is independent of projectile or target.
 
 The problem of how to include the running coupling in the dipole
 evolution kernel ${\mathcal K}(r, r^\prime)$ has been studied in
 ref.~\cite{ARS1}.  The form of the kernel with running coupling is
 fixed by the following two requirements.  The first is that in the
 emission probability of a soft gluon with transverse momentum
 ${\mathbf k}$ the coupling runs as $\alpha_s({\mathbf k}^2)$.  The
 second is that the emission probability in impact parameter space
 should generate the same singularities in Borel space as the Borel
 transformed emission probability in momentum space.  Then the form of
 the one-dipole emission probability $\Phi^{(1)}(z, r)$ is
 completely determined.  Explicitly it is 
 \begin{eqnarray} 
 \Phi^{(1)}(z, r) &=&
 \frac{C_F}{\pi^3} \,  \Phi^{(0)}(z,r) \,
 \int^z \frac{\d z_1}{z_1} \, 
 \int \d^2 {\mathbf r}_1 \int_0^\infty \d
 \tau \, \d \hat{\tau} J_1(\tau) J_1(\hat{\tau}) \, \int_0^1 \frac{\d
   \omega}{\omega^{1/2} (1-\omega)^{1/2}}
 \nonumber \\
 &\quad& \times \frac{1}{r_1^2 \, \hat{r}_1^2} \left\{ \alpha_s\left(
     \frac{\lambda^2}{R_1^2} \right) r^2 + \left[
     \alpha_s\left(\frac{\lambda^2}{r_1^2} \right) -\alpha_s \left(
       \frac{\lambda^2}{ R_1^2} \right) \right] \hat{r}_1^2 \right.
 \nonumber \\
 &\quad& \quad \quad \quad \quad  
     \left. + \left[
     \alpha_s\left(\frac{\lambda^2}{\hat{r}_1^2}\right) -\alpha_s
     \left( \frac{\lambda^2}{R_1^2} \right) \right] r_1^2 \right\}
 \, .
 \label{phirun}
 \end{eqnarray} 
 where 
 \begin{equation} \lambda^2(\tau, \hat{\tau}, \omega) =
 (\tau^2)^\omega (\hat{\tau}^2)^{(1-\omega)} \, , \hspace{.5cm}
 R_1^2(\omega) = (r_1^2)^\omega (\hat{r}_1^2)^{(1-\omega)} \, .
 \label{scaledefs}
 \end{equation} 
 In the limit of fixed $\alpha_s$ the above equation reproduces
 the expression for the one-dipole emission $\Phi^{(1)}$ of
 eq.~(\ref{Phi1}).  It turns out that in impact parameter space the
 scale of the coupling is not a simple function of the distances
 $r_1$, $\hat{r}_1$ but is weighted with three further parameters
 $\tau$, $\hat{\tau}$ and $\omega$.
 
 The Borel image $\tilde{\Phi}^{(1)}(z,r;b)$ is defined, as usual,
 by the transformation 
 \begin{equation} \Phi^{(1)}(z, r) = \int_0^\infty
 {\mathrm d} b \, \tilde{\Phi}^{(1)}(z, r; b) \, e^{-b /
   \alpha_s(Q^2)} \, .
 \label{bortransphi}
 \end{equation} 
 This transformation can be easily inverted when the one-loop
 running coupling is used.  Then the Borel image
 $\tilde{\alpha}_s({\mathbf k}^2 / Q^2; b)$ of $\alpha_s({\mathbf
   k}^2)$ is 
 \begin{equation} \tilde{\alpha}_s({{\mathbf k}^2 \over Q^2}; b) =
 \left( \frac{{\mathbf k}^2}{Q^2} \right)^{-b \beta_0} \, ,
 \hspace{1cm} \beta_0 = \frac{1}{4 \pi} \left( \frac{11}{3} N_c -
   \frac{2}{3} N_f \right) \, ,
 \label{borelalpha}
 \end{equation} 
 and from eqs.~(\ref{phirun},\ref{bortransphi}) we obtain 
 \begin{eqnarray}
 \tilde{\Phi}^{(1)}(z, r; b) &=& \frac{C_F}{\pi^3} \, 
 \Phi^{(0)}(z, r) \, \ln(z) \,
 \int \d^2 {\mathbf r}_1 \int_0^\infty \d \tau \d
 \hat{\tau} J_1(\tau) J_1(\hat{\tau}) \int_0^1 \frac{\d
   \omega}{\omega^{1/2}\, (1-\omega)^{1/2}}
 \nonumber \\
 &\quad& \times \frac{1}{\tau^2}^{\omega b \beta_0}
 \frac{1}{\hat{\tau}^2}^{(1-\omega) b \beta_0} \frac{ (Q^2)^{b
     \beta_0}}{r_1^2 \, \hat{r}_1^2}
 \nonumber \\
 &\quad& \times \left[ (r_1^2)^{b \beta_0} \hat{r}_1^2
   +(\hat{r}_1^2)^{b \beta_0} r_1^2 +2 {\mathbf r}_1 \cdot
   \hat{{\mathbf r}}_1 (r_1^2)^{\omega b \beta_0}
   (\hat{r}_1^2)^{(1-\omega) b \beta_0} \right]  \, .
 \label{borelphi}
 \end{eqnarray} 
 The integrals over $z_1, {\mathbf r}_1$, $\tau$, $\hat{\tau}$
 and $\omega$ can be performed and the result is 
 \begin{equation}
 \tilde{\Phi}^{(1)}(z, r; b) = -\frac{2 C_F}{\pi} \ln(z)
 \frac{\Gamma(-b \beta_0)} {\Gamma(1+ b\beta_0)} \left( \frac{Q^2
     r^2}{4} \right)^{b \beta_0} \Phi^{(0)}(z, r) \, .
 \label{renormalon}
 \end{equation} 
 From eqs.~(\ref{borelphi}, \ref{renormalon}) we can construct
 the dipole evolution kernel in Borel space $\tilde{\mathcal K}(r,
 r^\prime; b)$.  It consists of two pieces.  The virtual contribution
 is the coefficient of the $\Phi^{(0)}$ in eq.~(\ref{renormalon}) with
 the sign inverted as required by unitarity \cite{Mueller}.  The real
 contribution is obtained by the coefficient of $\Phi^{(0)}$ in
 eq.~(\ref{borelphi}).  After renaming $r_1 \rightarrow r^\prime$,
 changing integration variables as in eq.~(\ref{chvars}) and
 performing the integration over $\hat{r}^\prime$ we obtain the
 explicit form of the kernel. It is\footnote{The factor of 2 in front of 
 the second term of this equation has been inadvertently omitted in 
 eq.~(16) of ref. \cite{ARS1}.}
 \begin{eqnarray} \tilde{\mathcal K}(r,
 r^\prime; b) &=& \frac{N_c}{\pi} \left( \frac{Q^2 r^2}{4} \right)^{b
   \beta_0} \frac{ \Gamma(-b \beta_0)}{\Gamma(1+ b \beta_0)}
 \delta(r-r^\prime)
 \nonumber \\
 &+& 2 \frac{N_c}{\pi^2} \, \frac{1}{r^\prime} \left( \frac{Q^2
     r^{\prime 2}}{4} \right)^{b \beta_0} \int_0^1 \frac{\d
   \omega}{\omega^{1/2} (1-\omega)^{1/2}} \frac{ \Gamma(1-\omega b
   \beta_0) \, \Gamma(1-(1-\omega) b \beta_0)} { \Gamma(1+\omega b
   \beta_0) \, \Gamma(1+(1-\omega) b \beta_0)}
 \nonumber \\
 &\times& \left\{ \left( \frac{ r_>^2}{r^{\prime 2}} \right)^{b
     \beta_0-1} {_2F_1} \left(1- b \beta_0, 1-b \beta_0;1;
     \frac{r_<^2}{r_>^2}\right) \right.
 \nonumber \\
 &\quad& \quad +\left( \frac{r^2-r^{\prime2}}{ r_>^2} \right) \left( \frac{
     r_>^2}{r^{\prime 2}} \right)^{\omega b \beta_0} {_2F_1} \left(1-
   \omega b \beta_0, 1-\omega b \beta_0;1; \frac{r_<^2}{r_>^2}\right)
 \nonumber \\
 &\quad& \quad -\left.  \left( \frac{r_>^2}{r^{\prime 2}} \right)^{\omega b
     \beta_0} {_2F_1} \left( -\omega b \beta_0, -\omega b \beta_0;1;
     \frac{r_<^2}{r_>^2}\right) \right\} \, ,
 \label{runker}
 \end{eqnarray} 
  where $r_< = {\mathrm min}(r, r^\prime)$ and 
 $ r_> = {\mathrm max}(r, r^\prime)$.  Finally, the evolution equation for 
 the dipole density in Borel space reads 
 \begin{equation} \frac{\partial}{\partial Y}
 \tilde{n}(Y, r, \rho; b) = \int_0^\infty \d r^\prime \int_0^b \d
 b^\prime \tilde{\mathcal K}(r, r^\prime; b^\prime) \, \tilde{n}(Y,
 r^\prime, \rho; b-b^\prime) \, ,
 \label{dipevol} 
 \end{equation} with boundary condition 
 \begin{equation} \tilde{n}(Y=0, r, \rho; b) = r \,
 \delta(r-\rho) \, \delta(b) \, .
 \label{bc}
 \end{equation}

 \section{ The Borel singularity structure of the dipole cascade}
 
 To study the Borel singularity structure generated by the dipole
 kernel $\tilde{\mathcal K}$ we consider its action on a test function
 of the form $(r^2)^\gamma$.  Defining the function $\chi(\gamma, b)$ by 
 \begin{equation} \int_0^\infty \d r^\prime \, \tilde{\mathcal K}(r, r^\prime;
 b) (r^{\prime 2})^\gamma = \frac{N_c}{\pi} \chi(\gamma, b)
 \left(\frac{Q^2 r^2}{4}\right)^{b \beta_0} (r^2)^{\gamma} \, ,
 \label{chidef}
 \end{equation} 
 we obtain the expression of $\chi(\gamma, b)$ from
 eq.~(\ref{runker}), 
 \begin{eqnarray} \chi(\gamma, b) &=& \frac{\Gamma(-b
   \beta_0)}{\Gamma(1+ b \beta_0)}
 \nonumber \\
 &+& \frac{\Gamma(-\gamma-b \beta_0)}{\Gamma(1+ \gamma +b \beta_0)}
 \frac{1}{\pi} \int_0^1 \frac{\d
   \omega}{\omega^{1/2}{(1-\omega)}^{1/2}} \frac{\Gamma(1-\omega b
   \beta_0)}{\Gamma(1+\omega b \beta_0)} \frac{\Gamma(1-(1-\omega) b
   \beta_0)}{\Gamma(1+(1-\omega) b \beta_0)}
 \nonumber \\
 &\times& \left[ \frac{\Gamma(1+\gamma)}{\Gamma(-\gamma)}
   \frac{\Gamma(b \beta_0)}{\Gamma(1 - b \beta_0)} -2 \,
  \frac{\Gamma(1+\gamma +(1-\omega) b \beta_0)} {\Gamma(1-\gamma
     -(1-\omega) b \beta_0)}
   \frac{\Gamma(1+\omega b \beta_0)}{\Gamma(1-\omega b \beta_0)}
  \right] \, ,
 \label{chiform}
 \end{eqnarray} 
 where the first term comes from virtual correction and the
 second from real emission.  The virtual contribution to $\chi(\gamma,
 b)$ contains a series of poles at $b \beta_0 = 1, \, 2, \, 3 \, ...$
 which are identified with the IR renormalons and correspond to power
 corrections of ${\mathcal O}( (m_N^2/Q^2)^n)$, $ n=1,2,...$ Note that
 these poles are independent of the specific form of the test
 function.  This set of poles, resulting from the exponentiation of
 soft radiation, has also been derived in the context of the Drell-Yan
 process in ref. \cite{StermKorch}.  In addition, we observe the
 presence of a series of poles at $b \beta_0 = n-\gamma$, $n=0,1,2
 ...$ generated by the $\Gamma(-\gamma-b \beta_0)$ dependence of the
 real contribution to $\chi(\gamma, b)$.  For $Re(\gamma) \ge m$ these
 poles correspond to IR renormalons for $n>m$.  Their IR origin is
 established by observing that these singularities arise from the
 $r^\prime > r$ integration region of eq.~(\ref{dipevol}), where the
 offspring dipole is emitted with size larger than the parent dipole.
 Taken at face value the $\gamma$-dependent poles indicate the
 presence of ${\mathcal O}( (m_N^2/Q^2)^{n-\gamma})$ power
 corrections.  For the moment it suffices to note that in this section
 we are studying the dipole cascade independently of its embedding in
 a particular physical process.  The interpretation of the
 $\gamma$-dependent power corrections in terms of OPE matrix elements
 is not an issue here.
 
 Inspecting the evolution equation (\ref{dipevol}) we now examine
 whether the kernel introduces singularity at the lower end of the
 $b^\prime$-integration.  The answer is that for fixed $r$ the limit
 $b \rightarrow 0$ of $\tilde{{\mathcal K}}(r, r^\prime, b)$ is not
 singular.  The $b=0$ pole appearing in the virtual contribution to
 $\tilde{\mathcal K}$ is of UV origin and cancels against a
 corresponding singularity in the real contribution.  The presence of
 such a singularity can be seen already in the fixed coupling case.
 The integral in eq.~(\ref{Phi1}) is logarithmically divergent in the
 region $r_1 \rightarrow 0$.  One way of tracing the cancellation of
 the UV singularities is to impose a lower cutoff in the impact
 parameter integration as in ref.~\cite{Mueller}.  Alternatively, in
 the case of running coupling we can use the Borel parameter $b$ as a
 dimensional regulator.  Defining \begin{equation} \zeta = \frac{ r_<^2}{ r_>^2}
 \label{zeta}
 \end{equation} we note that the UV divergences in the real part of
 $\tilde{\mathcal K}$ arise from the region $\zeta \rightarrow 1$.
 This is the region where the gluon is emitted at impact parameter
 almost equal to that of the parent dipole. In the limit $b
 \rightarrow 0$ the only UV divergent contribution comes from the
 first term inside the curly brackets of eq.~(\ref{runker}). It is due
 to the divergence of the hypergeometric function ${_2F_1}(1-b
 \beta_0, 1-b \beta_0;1;\zeta=1)$ at $b \rightarrow 0$.  To
 parametrise this divergence we use the linear transformation
 \cite{Erdelyi} \begin{equation} {_2F_1}(1-b \beta_0, 1-b \beta_0; 1; \zeta) = (1-
 \zeta)^{-1+2 b \beta_0} \, {_2F_1}(b \beta_0, b \beta_0;1;\zeta) \, .
 \label{hyptrans}
 \end{equation} Then, in the limit $b \rightarrow 0$ the kernel takes the form
 \begin{eqnarray} \tilde{\mathcal K}(r, r^\prime; b) &=& \frac{N_c}{\pi} \left(
   \frac{Q^2 r^2}{4} \right)^{b \beta_0} \left\{ \frac{1}{(-b
     \beta_0)} \delta(r- r^\prime) \right.
 \nonumber \\
 &\quad & \left.  +\frac{2}{ r^\prime} \, \frac{1}{\pi} \, \int_0^1
   \frac{\d \omega}{\omega^{1/2}(1-\omega)^{1/2}} (1-\zeta)^{-1+ 2 b
     \beta_0} \, {_2F_1}(0, 0; 1; \zeta) + ({\mathrm UV \, regular})
 \right\} \, .
 \label{kerzeta}
 \end{eqnarray} The singular factor $(1-\zeta)^{-1+ 2 b \beta_0}$ can be
 expressed in terms of distributions as \begin{equation} (1-\zeta)^{-1+ 2 b
   \beta_0} = \frac{1}{2 b \beta_0} \, \delta(1- \zeta) + \left[
   \frac{1}{1-\zeta} \right]_+ + ...
 \label{plusdist}
 \end{equation} Substituting this into eq.~(\ref{kerzeta}) it follows that the
 UV singular term in the virtual part of $\tilde{\mathcal K}$ cancels
 against the one in the real part.  Since the kernel is regular around
 $b=0$ we compute the power series expansion in $b$ of $\chi(\gamma,
 b)$. From eq.~(\ref{chiform}) we obtain \begin{equation} \chi(\gamma, b) =
 \chi(\gamma) +b \beta_0 \chi^{(1)}(\gamma) + {\mathcal O}(b^2
 \beta_0^2) \, .
 \label{chiexpand}
 \end{equation} where \begin{equation} \chi^{(1)}(\gamma) = -\frac{1}{\gamma} \chi(\gamma)
 -2 \Psi(1) \chi(\gamma) +\frac{1}{2} \chi(\gamma)^2 +\frac{1}{2}
 \chi^\prime(\gamma)
 \label{chi1}
 \end{equation} The ${\mathcal O}(1/b)$ singular terms have cancelled as
 anticipated and the ${\mathcal O}(b^0)$ term is the BFKL spectral
 function $\chi(\gamma)$, eq.~(\ref{chifn}).
 
 Returning to the dipole evolution equation (\ref{dipevol}) we
 introduce the usual Laplace transform with respect to the rapidity
 $Y$ \begin{equation} \tilde{n}_{\omega}(r, \rho; b) = \int_0^\infty \d Y
 e^{-\omega Y} \tilde{n}(Y, r, \rho; b) \, ,
 \label{LaplY}
 \end{equation} and the anomalous dimension $\gamma$ via the Mellin transform
 \begin{equation} \tilde{n}_{\omega, \gamma}(b) = \int_0^\infty \frac{\d r^2}{r^2}
 \left( \frac{r^2}{\rho^2} \right)^{-\gamma} \tilde{n}_\omega(r, \rho;
 b) \, .
 \label{anomdim}
 \end{equation} Then eq.~(\ref{dipevol}) with the boundary condition (\ref{bc})
 through the use of (\ref{chidef}) becomes an equation for the
 spectral amplitudes $\tilde{n}_{\omega, \gamma}(b)$, \begin{equation} \omega \,
 \tilde{n}_{\omega, \gamma}(b) = 2 \delta(b) + \frac{N_c}{\pi}
 \int_0^b \d b^\prime \, \chi(\gamma-b^\prime \beta_0, b^\prime) \,
 \left( \frac{Q^2 \rho^2}{4} \right)^{b^\prime \beta_0} \,
 \tilde{n}_{\omega, \gamma-b^\prime \beta_0}(b-b^\prime) \, .
 \label{volterra}
 \end{equation} This equation is universal for all semihard processes.  The
 scale $Q^2$ is only constrained by the requirement $Q^2 \gg
 \Lambda_{\mathrm QCD}^2$ and its specific value is otherwise
 arbitrary.  It is identified with the $\gamma^\star$ virtuality here
 because we consider the specific case of small-$x$ D.I.S.  The dipole
 density is reconstructed from the spectral amplitudes as \begin{equation}
 \tilde{n}(Y, r, \rho; b) = \int \frac{\d \omega}{2 \pi i} e^{\omega
   Y} \int \frac{ \d \gamma}{2 \pi i} \left( \frac{r^2}{\rho^2}
 \right)^\gamma \tilde{n}_{\omega, \gamma}(b) \, .
 \label{inverse}
 \end{equation} Eq.~(\ref{volterra}) is an integral equation of the Volterra
 type.  Such equations are known not to have eigenfunctions for
 bounded kernels.  One way of constructing solutions is by iteration.
 In our case, the iterative solution is given by the formal expression
 \begin{equation} \tilde{n}_{\omega, \gamma}(b) = \sum_{k=0}^{\infty}
 \tilde{n}_{\omega, \gamma}^{(k)}(b) \, ,
 \label{itseries}
 \end{equation} with \begin{equation} \tilde{n}_{\omega, \gamma}^{(0)}(b) = \frac{2}{\omega}
 \delta(b) \, , \hspace{1cm} \tilde{n}_{\omega, \gamma}^{(1)}(b) =
 \frac{2}{\omega} \left( \frac{N_c}{\pi \omega} \right) \left(
   \frac{Q^2 \rho^2}{4} \right) ^{b \beta_0} \, \chi(\gamma-b \beta_0,
 b) \, ,
 \label{init}
 \end{equation} 
 and
 \begin{eqnarray} \tilde{n}_{\omega, \gamma}^{(k)}(b) &=& \frac{2}{\omega} \left(
   \frac{N_c}{ \pi \omega } \right)^k \, \left( \frac{Q^2 \rho^2}{4}
 \right)^{b \beta_0} \, \int_0^b \d b_{k-1} \, \chi(\gamma-(b-b_{k-1})
 \beta_0, b-b_{k-1})
 \nonumber \\
 &\quad& \times \int_0^{b_{k-1}} \d b_{k-2} \, \chi(\gamma - (b-
 b_{k-2}) \beta_0, b_{k-1}-b_{k-2}) \, ...
 \nonumber \\
 &\quad& \times \int_0^{b_3} \d b_2 \, \chi(\gamma- (b-b_2) \beta_0,
 b_3-b_2)
 \nonumber \\
 &\quad& \times \int_0^{b_2} \d b_1 \, \chi(\gamma -(b-b_1)\beta_0,
 b_2-b_1) \, \chi(\gamma -b \beta_0, b_1) \, ,
 \label{formsol}
 \end{eqnarray} for $k \ge 2$.  The index $k$ counts the number of iterations
 of the kernel in eq.~(\ref{volterra}).
 
 Having identified the IR renormalon poles of $\chi(\gamma, b)$,
 eq.~(\ref{chiform}), we can study the Borel singularities of the
 spectral amplitude $\tilde{n}_{\omega, \gamma}(b)$ from
 eq.~(\ref{formsol}).  The contribution to $\tilde{n}^{(k)}_{\omega,
 \gamma}(b)$ from the virtual ($\gamma$-independent) terms only is
 \begin{eqnarray} 
 \frac{2}{\omega} \left( \frac{N_c}{ \omega \pi} \right)^k \,
 \left( \frac{Q^2 \rho^2}{4} \right)^{b \beta_0} &\times& \int_0^b \d
 b_{k-1} \, \frac{ \Gamma( -(b-b_{k-1})\beta_0)}{\Gamma(1+ (b-b_{k-1})
   \beta_0)} \int_0^{b_{k-1}} \d b_{k-2} \, \frac{\Gamma(
   -(b_{k-1}-b_{k-2}) \beta_0)} {\Gamma(1+ (b_{k-1}+b_{k-2}) \beta_0)}
 ...
 \nonumber \\
 &\times& \int_0^{b_2} \d b_1 \, \frac{ \Gamma(-(b_2-b_1)
   \beta_0)}{\Gamma(1+(b_2+b_1) \beta_0)} \frac{ \Gamma(-b_1\beta_0)}
 {\Gamma( 1+b_1\beta_0)} \, .
 \label{virtonly}
 \end{eqnarray} 
 Because the $b$-integrals are nested, for $b < 1/\beta_0$ no
 new singularity is introduced in the solution from these virtual
 corrections.  For $b> 1/\beta_0$ some of the $\Gamma$-function poles
 will be encountered on the $b$-integration paths turning the leading
 $b= 1/\beta_0$ pole into a branch cut.  Real contributions introduce
 singularities on the positive $b$-semiaxis for real $\gamma$, and for
 $0< \gamma <1$, the leading one is found from eqs.~(\ref{init},
 \ref{formsol}) to be at $b=\gamma/\beta_0$.  We note that for $b
 \beta_0< \gamma$ the real contributions to $\tilde{n}^{(k)}_{\omega,
   \gamma}(b)$ will not generate singularities because none of the
 poles of the $\chi$-functions in eq.~(\ref{formsol}) are encountered
 along the nested $b$-integration paths.  For $b \beta_0>\gamma$, as
 in the case of virtual contributions, poles are found along some of
 the $b$-paths turning the leading $b=\gamma/ \beta_0$ pole into a
 branch cut.  
 From eqs.~(\ref{chiform}, \ref{formsol}) it follows that
 there are also singularities for negative values of $b$.  This means
 that $b=\gamma/\beta_0$ is the position of the leading singularity
 for positive $b$ {\it only} when $\gamma$ is in the range
 $0<\gamma<1$.  If we were to take $\gamma$ in the range
 $n<\gamma<n+1,\ n \in Z$, then the leading IR singularity would be at
 $\gamma-n$, and the singularities for negative $b$ would be displaced further
 to the left on the negative $b$-semiaxis. 
 Although these singularities do not affect the estimate of the power 
 corrections from the non-perturbative effects, they reflect the fact that 
 as $\gamma$ increases the dipole densities generated from these test 
 functions  become increasingly IR divergent.
 
 The above argument can be repeated for the contributions to the
 solution from cross products of virtual terms for some of the
 $\chi$-functions and real terms for the rest.  The net result of
 this analysis is that the region of analyticity in Borel space of
 $\tilde{n}_{\omega, \gamma}(b)$ with $0<\gamma<1$ contains the
 interval $0 < b < \gamma /\beta_0$ and the leading singularity is a
 branch cut at $b=\gamma /\beta_0$.  In the asymptotic limit $Y
 \rightarrow \infty$ the anomalous dimension is $\gamma = 1/2+ i \nu$
 with $\nu$ the spectral parameter to be integrated over in
 eq.~(\ref{inverse}).  Such a $Re(\gamma)$ implies the presence of
 power corrections of ${\mathcal O}(m_N/Q)$ for the dipole density
 $n(Y, r, \rho)$.  These $1/Q$ power corrections have also been
 reported in the context of conventional BFKL approach in
 ref.~\cite{Levin}.
 
 The evolution equation (\ref{dipevol}) is not conformally invariant
 because the action of the dipole kernel results in a shift of
 $\gamma$ by $b^\prime \beta_0$ in eq.~(\ref{volterra}).  However,
 having established above a region of analyticity in Borel space that
 contains $b=0$ we can use the $b$-expansion as a measure of deviation
 from the conformal limit.  From eqs.~(\ref{chiexpand}, \ref{formsol})
 the $b$-expansion of the spectral amplitude generated after $k$
 iterations of the kernel is 
 \begin{eqnarray} \tilde{n}_{\omega, \gamma}^{(k)}(b)
 &=& \frac{2}{\omega} \left( \frac{N_c}{\pi \omega } \right)^k \left\{
   \frac{b^{k-1}}{(k-1)!} \chi(\gamma)^k + \frac{b^k}{(k-1)!}
   \left[\beta_0 \ln \left(\frac{Q^2 \rho^2}{4} \right) \chi(\gamma)^k
     + \beta_0 \chi^{(1)}(\gamma) \chi(\gamma)^{k-1} \right] \right.
 \nonumber \\
 &\quad& \hspace{2cm} \left.  -\frac{b^k}{2} \frac{(k+1)}{(k-1)!}
   \beta_0 \chi(\gamma)^{k-1} \frac{\d \chi(\gamma)}{\d \gamma}
   +{\mathcal O}(b^{k+1}) \right\} \, .
 \label{bexp}
 \end{eqnarray} 
 The full spectral amplitude $n_{\omega, \gamma}(b)$, defined in
 eq.~(\ref{itseries}), can be constructed in the small-$b$ region from
 summing the powers of $b$.  Summation of the leading powers of $b$,
 i.e. retaining only the ${\mathcal O}(b^{k-1})$ terms in
 eq.~(\ref{bexp}) is required in the region 
 \begin{equation} b \ll 1, \hspace{1cm}
 \frac{N_c \chi(\gamma)}{\pi \omega } b \sim 1 \, .
 \label{resum_region}
 \end{equation} 
 This `resummation' of the ${\mathcal O}( N_c \chi(\gamma) b/
 (\omega \pi))$ terms is well defined if the ${\mathcal O}(b^k)$ terms
 are small relative to the ${\mathcal O}(b^{k-1})$ terms in
 eq.~(\ref{bexp}).  For the ${\mathcal O}( b^k \ln (Q^2 \rho^2/4))$
 term this translates to the requirement 
 \begin{equation} b \beta_0 \ln \frac{Q^2
   \rho^2}{4} \ll 1 \, \Leftrightarrow b \left(
   \frac{1}{\alpha_s(Q^2)} - \frac{1}{\alpha_s( 1/ \rho^2)} \right)
 \ll 1 \, .
 \label{conf_region}
 \end{equation} 
 Hence, leading $b$-power resummation is valid in the conformal
 limit of fixed coupling.  In this case the Borel transform can be
 readily inverted in terms of a fixed $\alpha_s$ to yield the well
 known answer for the BFKL spectral amplitude 
 \begin{equation} n_{\omega, \gamma}
 = \sum_{k=0}^{\infty} \frac{2}{\omega} \left[ 1 + \frac{N_c \alpha_s
     \chi(\gamma)}{\pi \omega } + \left( \frac{N_c \alpha_s
       \chi(\gamma)}{\pi \omega } \right)^2 + ...  \right] =
 \frac{2}{\omega - N_c \alpha_s \chi(\gamma)/\pi} \, ,
 \label{conflim}
 \end{equation} 
 and via eq.~(\ref{inverse}) the asymptotic solution for the
 dipole density is obtained \cite{MuellPatel} 
 \begin{equation} n(r, \rho, Y) =
 \frac{1}{2} \frac{r}{\rho} \, \frac{\exp[ (\alpha_P-1)Y]}{
   \sqrt{(7/2) \alpha_s N_c \zeta(3) Y}} \, \exp \left( -\frac{\pi
     \ln^2(r/\rho)}{14 \alpha_s N_c \zeta(3) Y} \right) \, ,
 \label{BFKLpom}
 \end{equation} 
 with $\alpha_P-1 = 4 (\alpha_s N_c/ \pi) \ln2$.

 \section{ Small-$x$ D.I.S. and power corrections}
 
 The structure functions in the small-$x$ region can be written in the
 following form \begin{equation} F_{T,L}(x, Q^2) = \frac{Q^2}{4 \pi \alpha_{em}}
 \int_0^1 \d z \, \int \d^2 {\mathbf r} \, \Phi^{(0)}_{T,L}(z, r)
 \,\sigma_{d,N}(Y=\ln(z/x), r) \, .
 \label{struct0}
 \end{equation} Using the definition of the dipole-nucleon cross section,
 eq.~(\ref{sigman}), the above expression becomes \begin{eqnarray} F_{T,L}(x,
 Q^2) &=& \frac{Q^2}{4 \pi \alpha_{em}} \int_0^1 \d z \int
 \d^2{\mathbf r} \, \Phi^{(0)}_{T,L}(z, r) \int \frac{\d^2\rho}{2
   \pi \rho^2} \, n(Y, r, \rho) \, \sigma_0(\rho, m_N)
 \nonumber \\
 &=& \frac{Q^2}{4 \pi \alpha_{em}} \int \frac{\d^2\rho}{2 \pi \rho^2}
 \, N_{T,L}(Y, \rho) \, \sigma_0(\rho, m_N) \, .
 \label{struct1}
 \end{eqnarray} 
 This equation is the factorisation theorem for the structure
 function in the small-$x$ regime.  As mentioned in the introduction, it
 contains in addition to convolution in the longitudinal momentum
 fraction a convolution in impact parameter.  $\Phi^{(0)}$ is the
 lowest order transition probability for $\gamma^\star \rightarrow q
 \bar{q}$.  (This is the lowest order impact factor of
 ref.~\cite{Balitsky}.)  It can be calculated reliably in perturbation
 theory in the limit $Q^2 \gg \Lambda^2_{\mathrm QCD}$.  The dipole density
 $n(Y, r, \rho)$ contains the universal evolution of the initial 
 $q$-$\bar{q}$ dipole into a cascade of dipoles with ordered rapidities.
 This is also calculable in perturbation theory as we saw in the
 previous section.  Finally $\sigma_{0}(\rho, m_N)$ is the cross
 section for the absorption of a dipole at impact parameter $\rho$ by
 the nuclear target.  This is a non-perturbative quantity that
 normalises the structure function and introduces dependence on the 
 mass scale characteristic of the nucleon target.

 So far we have derived the evolution equation that determines the
 dipole density $n(Y, r, \rho)$.  For completeness let us briefly
 review the calculation of $\Phi^{(0)}$ defined in eq.~(\ref{Phidef}).
 In momentum space and to ${\mathcal O}(\alpha_s^0)$ the
 $\gamma^\star$ wave function is \cite{lightfront} \begin{equation}
 \psi^{(0)}_{T,L}(z, {\mathbf k}) = \frac{1}{2 q^+ \sqrt{z(1-z)}} \,
 \int_{-\infty}^\infty \frac{\d k^-}{2 \pi} \, \overline{u} \gamma^+
 G_\mu(k, q) \gamma^+ v \, \epsilon_{T,L}^\mu(q) \, ,
 \label{psi0}
 \end{equation} where $u$ and $v$ are the spinors for the outgoing fermions,
 $\epsilon_{T,L}^{\mu}(q)$ is the polarisation vector of
 $\gamma^\star$ and $G_\mu$ is the $\gamma^\star q \overline{q}$
 Green's function \begin{equation} G_\mu(k, q) = i e Z_q \frac{ \not{k} +
   m_q}{k^2-m_q^2 + i \varepsilon} \gamma_\mu \frac{( \not{k} - \not{
     \! q}) +m_q}{ (k-q)^2-m_q^2+i \varepsilon} \, ,
 \label{G}
 \end{equation} with $Z_q$ the electric charge of the quark.  Performing the
 numerator algebra and the $k^-$ integration we obtain \begin{eqnarray}
 \psi^{(0)}_T(z, {\mathbf k}) &=& -\frac{e Z_q}{2 q^+ \sqrt{z (1-z)}}
 \, \frac{ \overline{u}(z \not{ \! {\mathbf \epsilon}} \not{{\mathbf
       k}} -(1-z) \not{\! {\mathbf \epsilon}} \not{{\mathbf k}} -m_q
   \not{\! {\mathbf \epsilon}} ) \gamma^+ v} { {\mathbf k}^2 +
   \mu^2} \, ,
 \\
 \psi^{(0)}_L(z, {\mathbf k}) &=& -\frac{e Z_q}{2 q^+ \sqrt{z (1-z)}}
 \, z(1-z) Q \, \frac{\overline{u} \gamma^+ v}{{\mathbf
     k}^2+\mu^2} \, ,
 \label{psimom}
 \end{eqnarray} where \begin{equation} \mu^2 = z(1-z) Q^2 +m_q^2 \, .
 \label{mu}
 \end{equation} 
 After taking the impact parameter transformation defined by \begin{equation}
 \psi^{(0)}_{T,L}(z, {\mathbf r}) = \int \frac{\d ^2 {\mathbf k}}{(2
   \pi)^2} \, e^{i {\mathbf k}{\mathbf r}} \psi^{(0)}_{T,L}(z,
 {\mathbf k}) \, ,
 \label{impdef}
 \end{equation} and squaring as in eq.~(\ref{Phidef}) we find \begin{eqnarray}
 \Phi^{(0)}_T(z,  r) &=& \frac{2 N_c e^2 Z_q^2}{(2 \pi)^2}
 \left\{ [z^2 + (1-z)^2] \mu^2 K_1(\mu r)^2 + m_q^2
   K_0(\mu r)^2 \right\} \, ,
 \nonumber \\
 \Phi^{(0)}_L(z, r) &=& \frac{8 N_c e^2 Z_q^2}{(2 \pi)^2}
 z^2(1-z)^2 Q^2 K_0(\mu r)^2 \, .
 \label{phi0}
 \end{eqnarray} 
 These expressions have also been derived in ref.~\cite{NikZak1}
 from the imaginary part of the $\gamma^\star g$ forward amplitude.
 
 The factorised expression for the structure functions $F_{T,L}$,
 eq.~(\ref{struct1}), allows us to study the effects of soft radiation
 in the small-$x$ region beyond the leading power of $Q$.  To this end
 we shall follow the standard lore of renormalon analysis
 \cite{renormalons}.  This means that we shall study the singularity
 structure of the Borel image $\tilde{F}_{T,L}(x, Q^2; b)$ defined in
 the usual way, 
 \begin{equation} F_{T,L}(x, Q^2) = \int_0^\infty \d b \,
 \tilde{F}(x, Q^2; b) \, e^{-b/ \alpha_s(Q^2)} \, .
 \label{borelF1}
 \end{equation} 
 We emphasise that even in the leading power in $Q$ analysis the
 structure functions always contain a non-perturbative component,
 denoted by $\sigma_0$ here.  However, we are looking for the Borel 
 singularities that are generated in the subasymptotic $Q$ regime by
 the components that are calculable in perturbation theory,
 $\Phi^{(0)}$ and $n$ in our case.  This is how perturbation theory is
 assumed to signal the presence of power corrections.  Inspecting
 eq.~(\ref{struct1}) we see that the source of Borel singularities is
 the dipole density $n$. $\Phi^{(0)}$ has no $\alpha_s$ dependence and 
 therefore it stays unaffected by the Borel transformation. 
 It affects the number density of the produced dipoles $N(Y, r, \rho)$ 
 by determining the distribution of the initial $q$-$\bar{q}$ dipole 
 in impact parameter (transverse size) space. 
 $\Phi^{(0)}$ contains infrared regulator set by the scale $\mu$, 
 eq.~(\ref{mu}). This follows from eqs.~(\ref{phi0}) and the asymptotic 
 behaviour of the Bessel-$K$  functions for $\mu r \rightarrow \infty$,
 \begin{equation}
 K_\nu(\mu r) \rightarrow \sqrt{\frac{\pi}{2 \mu r}} \, e^{-\mu r} \, 
 \left( 1+{\mathcal O} \left( \frac{1}{\mu r}\right) \right) \, .
 \label{asyK}
 \end{equation}
 Note that the leading term in this asymptotic expansion is independent 
 of the $\gamma^\star$ polarisation. 
 Since $\Phi^{(0)}(z,r)$ is dominated by values of $z$ away from the 
 end points $z=0$ and $z=1$, we can think of the initial 
 $\gamma^\star \rightarrow q \bar{q}$ fluctuation as having size 
 of ${\mathcal O}(1/Q)$. 
 The object that regulates the emission of dipoles of large sizes at the 
 end of the cascade is the dipole-nucleon cross section $\sigma_0$. 
 Indeed, dipoles of size $\rho \gg R_N$, with $R_N= {\mathcal O}(1/m_N)$ 
 the nucleon size, will not couple to the target.     
 From this discussion we can formulate the problem of the Borel singularities 
 of the small-$x$ structure functions as follows. 
 The Borel image of the structure functions is given by 
 \begin{equation}
 F_{T,L}(x, Q^2; b) = \frac{Q^2}{4 \pi \alpha_{em}} \, 
 \int_0^1 \d z \, \int \d^2 {\mathbf r} \, \Phi^{(0)}_{T,L}(z,r) \, 
 \tilde{\sigma}_{dN}(Y=\ln(z/x), r; b) 
 \label{borelsigma} 
 \end{equation}
 where $\tilde{\sigma}_{dN}$ is a solution to the evolution equation 
 \begin{equation}
 \frac{\partial}{\partial Y} \tilde{\sigma}_{dN}(Y, r; b) = 
 \int_0^\infty \d r^\prime \, \int_0^b \d b^\prime \, 
 \tilde{\mathcal K}(r, r^\prime; b^\prime) \, 
 \tilde{\sigma}_{d N}(Y, r^\prime; b-b^\prime) \, ,
 \label{evsigma}
 \end{equation}
 with kernel given in eq.~(\ref{runker}) and boundary condition 
 \begin{equation}
 \tilde{\sigma}_{d N}(Y=0, r) = \delta(b) \, \sigma_0(r, m_N) \, . 
 \label{sigmabc} 
 \end{equation}
  
 In the previous section we studied the action of the kernel
 $\tilde{{\cal K}}$ on the test functions $(r^2)^\gamma$, see
 eqs.~(\ref{chidef}, \ref{chiform}).  Then these functions where used
 as a basis for decomposing the dipole density in terms of the
 spectral amplitudes, eq.~(\ref{anomdim}), and we observed that this
 leads to $\gamma$-dependent IR renormalons.  If this procedure were
 applicable for constructing the structure functions $\tilde{F}_{T,L}$
 it would yield a leading IR renormalon at $b \beta_0=\gamma$, which
 could be to the left of unity for some $0 <\gamma<1$, corresponding
 to power corrections of ${\cal O}(1/ Q^p)$ with $p < 2$.  This would
 be in contradiction with the standard OPE expansion of the structure
 functions, which predicts leading power corrections of 
 ${\mathcal O}(1/Q^2)$, coming from the next-to-leading twist operators.
 
 In the case at hand such a contradiction does not arise for the
 following reason.  We have noted that IR renormalon singularities
 come from the integration region $r^\prime \gg r$ in
 eq.~(\ref{dipevol}), i.e. from emission at large impact parameters.
 This means that in eq.~(\ref{evsigma}) we are sampling large values
 of $r^\prime$. Although $\sigma_0(r^\prime, m_N)$ cannot be calculated 
 in pQCD it is controlled by a nucleon wavefunction which is suppressed 
 at large $r^\prime$, indicating that the probability to find a sufficiently 
 large primary dipole inside a nucleon is negligible.
 We can model $\sigma_0$ (as we shall do in the next section) by assuming 
 that it has the same asymptotic impact parameter dependence as that 
 of a virtual photon and mass scale $\mu \rightarrow m_N$. 
 Such a functional form, see eq.~(\ref{asyK}), cannot  
 be decomposed in the $(r^2)^\gamma$ basis.
 We emphasise that this does not mean that IR renormalons do not arise 
 from real emission. It means that the leading IR renormalon is anticipated 
 at $b \beta_0=1$ leading to ${\cal O}(1/Q^2)$ corrections consistent with
 the Wilson OPE expectation.  

 We conclude this section with the following remarks.
 The conventional (or Wilson) OPE
 expansion for large $Q$ cannot be used for small-$x$ resummation, as we
 noted in the introduction.  Semihard processes involve two large
 scales and a generalisation of the Wilson OPE is required.  Such an
 expansion would not only separate soft from hard particles but also
 high rapidity from low rapidity ones.  This type of formalism has
 been brought to a considerable degree of maturity by Balitsky in
 ref.~\cite{Balitsky}, although it is still far from the point of
 being a computational algorithm, like Wilson's OPE.  In the high
 energy OPE, evolution equations are non-linear beyond the LLA(x)
 approximation, whereas eq.~(\ref{dipevol}) is linear.  This is
 because our evolution equation does not resum the full set of the
 next-to-leading logarithms of $1/x$ but only a subset that involves
 the running of the coupling.  Hence, our solution is not expected to
 be unitary but it will parametrise sensitivity with respect to low
 transverse scales.

 \section{Numerical results}

 In this section we demonstrate by numerical means that the Borel transform 
 of a deeply inelastic scattering structure function is regular in the region 
 $0 < b\beta_0 < 1$ and has a singularity at $b\beta_0=1$.
 
 In perturbation theory we cannot calculate the cross section $\sigma_0$ of
 eq.~(\ref{sigmabc}) for  electron-proton scattering. 
 Nevertheless we know that it must be controlled by the nucleon wave function,
 which vanishes rapidly (assumed exponentially) for sufficiently 
 large impact parameter. 
 A reasonable model, therefore, is to take the impact factor dependence  
 calculated for the virtual photon, eq.~(\ref{phi0}). For the realistic case of
 D.I.S. this is indeed a model, whereas if we were considering
 the scattering of two virtual photons (onium-onium scattering) then
 this would be exact, up to a factor arising from the integration
 over the longitudinal momentum fraction, $z$, in the wavefunction of the
 target photon. 
 We require that the model nucleon wave function be normalisable and this 
 leads us to choose the case of a longitudinal virtual photon since for the 
 transverse photon the behaviour of the wave function for small impact 
 parameter gives rise to an ultraviolet divergence associated with the photon
 wave function renormalisation. The scale $\mu$ in eq.~(\ref{phi0})
 is set to a typical hadron scale of $\mu=m_N=1$~GeV. To this end we set 
 the Borel transform of $\sigma_{dN}^{(0)}(\rho)$ in 
 eqs.~(\ref{sigman}, \ref{sigmabc}) to
 \begin{equation} 
 \tilde{\sigma}_{dN}^{(0)}(\rho;b)=
 \rho^2  K_0(m_N \rho)^2 \delta(b).
 \label{d1} 
 \end{equation}
 We consider this modeling of the dipole-nucleon cross section to be 
 appropriate for studying the IR effects arising from emission and subsequent 
 interaction of large size dipoles. Note that, as stated in the previous 
 section, the large $r$ asymptotics of the $\gamma^\star$ wave function, which
 simulates the nucleon target here, is independent of polarisation. 
 So our choice of $K_0$ Bessel function in the previous equation is 
 convenient and plausible.
     
\begin{figure}
\hbox to \hsize{\hfill
\epsfxsize=0.46\hsize\epsffile{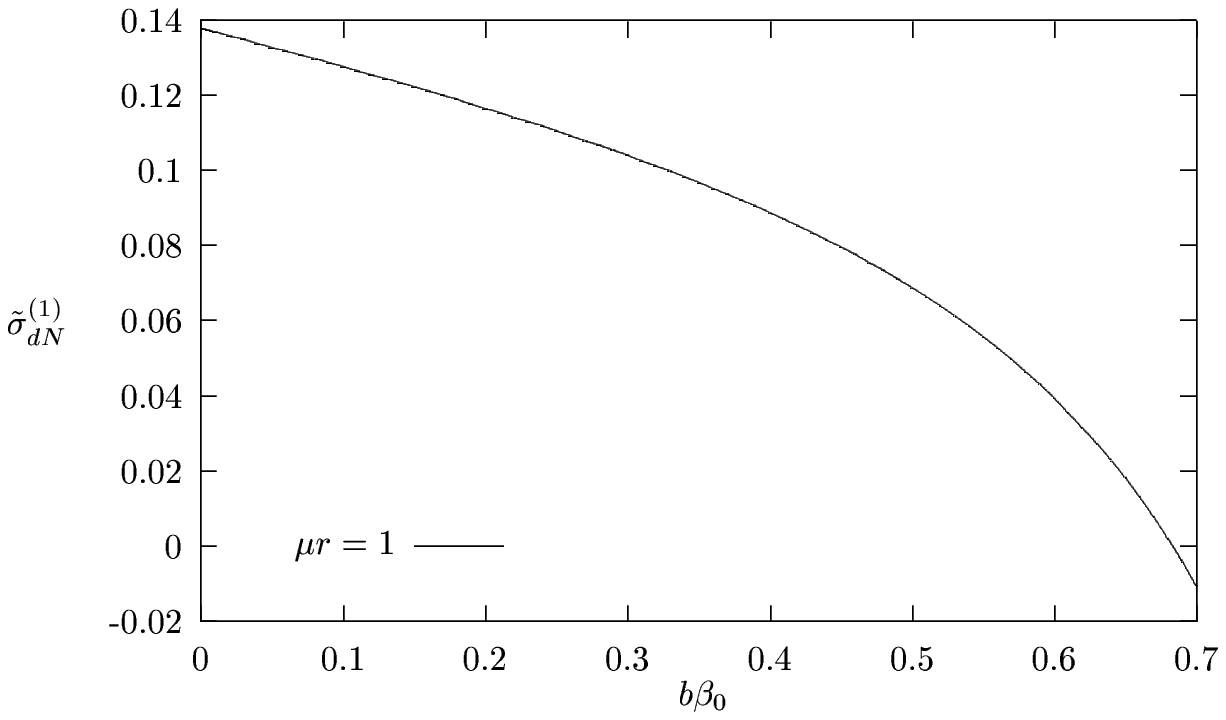}\hfill
\epsfxsize=0.46\hsize\epsffile{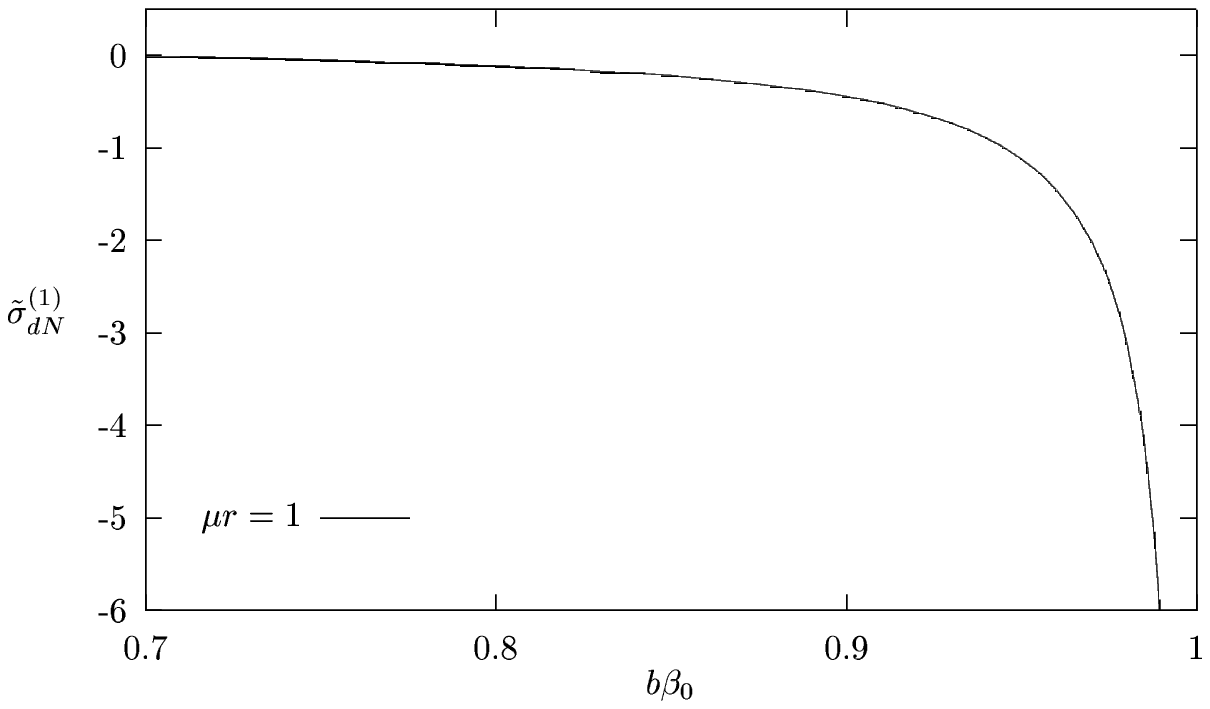}\hfill}
\caption[]{The Borel structure of
$  \tilde{\sigma}_{dN}^{(1)}(r;b)$, for
  $ \mu r =  1$. The graphs show intermediate $b \beta_0$ and $b
  \beta_0 \rightarrow 1$ behaviour.}
\label{conv_1_b}
\end{figure}

 The form of the input functions prevent an analytic solution  to  
 eq.~(\ref{evsigma}) from being found so we perform the integrations over
 $\omega$, $r^\prime$ and the convolution in $b^\prime$ numerically using 
 standard quadrature methods. More precisely we expand
\begin{equation}
\tilde{\sigma}_{dN}(Y,r;b)=\sum \frac{Y^n}{n!} 
\tilde{\sigma}_{dN}^{(n)}(r;b) \, ,
\label{d2} 
\end{equation}
so as to bring the evolution equation in the form
\begin{equation}
 \tilde{\sigma}_{dN}^{(n)}(r;b)=\int \d r^\prime \int_0^b \d b^\prime
 \tilde{\mathcal K}(r,r^\prime;b-b^\prime)
 \tilde{\sigma}_{dN}^{(n-1)}(r^\prime;b^\prime) \, . 
 \label{d3}
\end{equation} 
 The first convolution of the kernel with the longitudinal input function, 
 $\tilde{\sigma}_{dN}^{(1)}(r;b)$ exhibits the structure we expect from 
 section 4, see fig.~(\ref{conv_1_b}). 
 We observe that the leftmost singularity in the Borel plane appears at
 $b \beta_0 = 1$ and we identify this as the leading IR renormalon.
 There is no evidence of any singular behaviour in the region 
 $ 0 \leq b \beta_0 \leq 1$ which would have been in contradiction with
 the OPE expectation. No singularity emerges as $b \rightarrow 0$
 either, due to the cancellation between real and virtual parts of
 the kernel as discussed in section 3. On the other hand, the
 expected leading singularity at $ b \beta_0 = 1$ can be seen
 clearly from the right hand graph of fig.~(\ref{conv_1_b}). For the
 first iteration of the Borel transformed kernel, this singularity
 occurs as a pole and it is generated entirely by the virtual
 correction contribution to $\tilde{{\mathcal K}}$.
 
\begin{figure}
\hbox to\hsize{\hfill\epsfxsize=0.5\hsize\epsffile{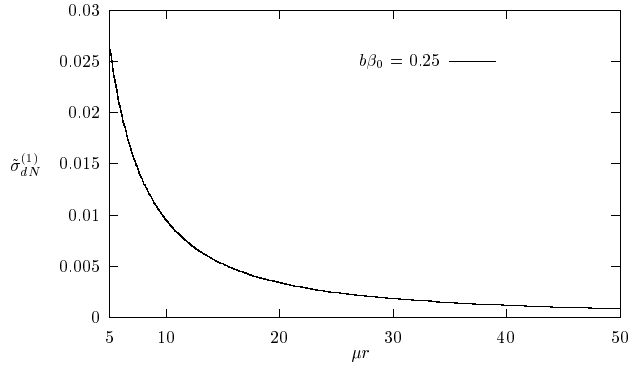}\hfill}
\caption[]{ The behaviour of $ \tilde{\sigma}_{dN}^{(1)}(r;b)$,
 in the intermediate $\mu r$ range and at $b \beta_0 = 0.25$, using
  the fit in eq.~(\ref{fititer1}).}
\label{conv_0_1_x}
\end{figure}

 There is a distinct difference between the asymptotic behaviour of
 $ \tilde{\sigma}_{dN}^{(0)}(r;b)$ and $ \tilde{\sigma}_{dN}^{(1)}(r;b)$ 
 as $\mu r \rightarrow \infty $, fig.~(\ref{conv_0_1_x}). 
 Whereas $ \tilde{\sigma}_{dN}^{(0)}(r;b)$ decays exponentially at large 
 $\mu r$, eq.~(\ref{asyK}), the asymptotic behaviour of
 $ \tilde{\sigma}_{dN}^{(1)}(r;b)$ is:
 \begin{equation}  
 \tilde{\sigma}_{dN}^{(1)}(r;b) 
 \rightarrow (\mu r)^{2(b \beta_0 -1)} \, .
 \label{asyiter1}
 \end{equation} 
 Examination of the kernel reveals that $\tilde{{\mathcal K}}$
 scales as $(\mu r^\prime)^{2b \beta_0-1}$ at large $ \mu
 r^\prime $ and this is the scaling behaviour which dominates in the
 convolution to determine the behaviour of $\tilde{\sigma}_{dN}^{(1)}$. For
 intermediate values of $\mu r$, the data obtained can be fitted
 accurately if we include logarithmic correction terms and we find
 \begin{equation}  
 \tilde{\sigma}_{dN}^{(1)}(r;b) \simeq (\mu r)^{2(b
 \beta_0 -1)}(a_1 + a_2\ln (\mu r) + a_3 \ln(\mu r)^2) \, ,
 \label{fititer1}
 \end{equation} 
 where $a_1$, $a_2$ and $a_3$ are fit parameters. This fit is exhibited in 
 fig.~(\ref{conv_0_1_x}) .
  
\begin{figure}
\vbox{%
\hbox to \hsize{\hfill
\epsfxsize=0.46\hsize\epsffile{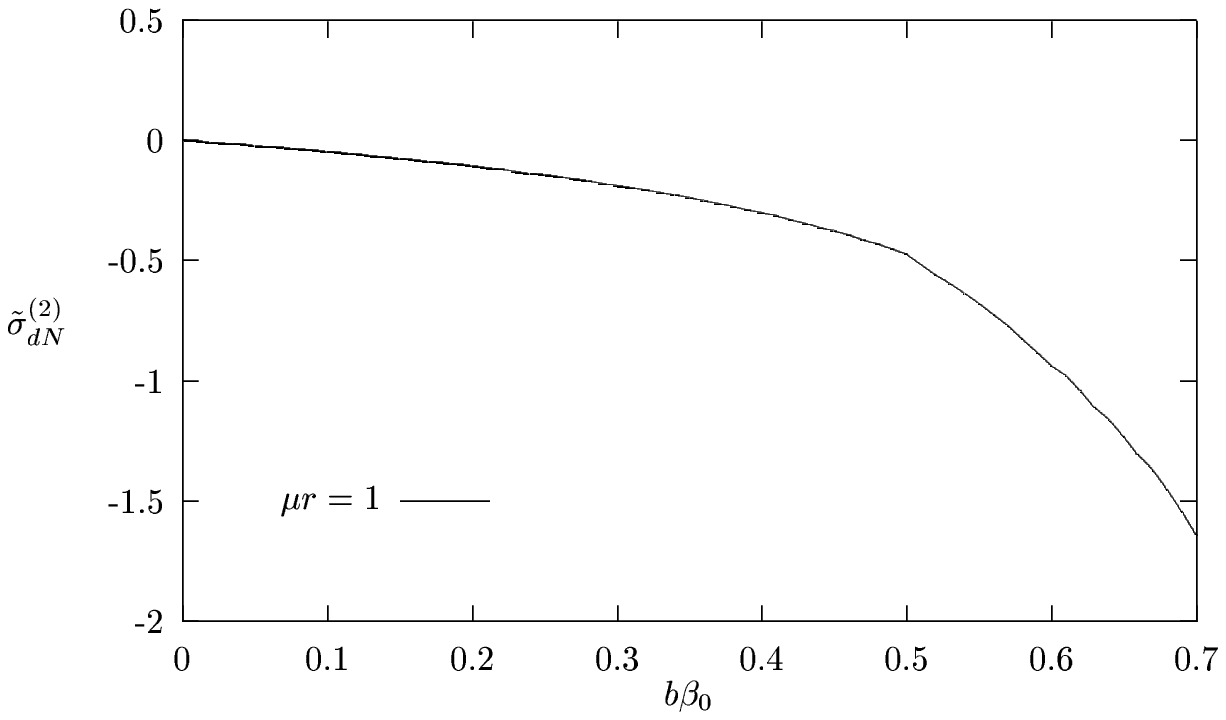}\hfill
\epsfxsize=0.46\hsize\epsffile{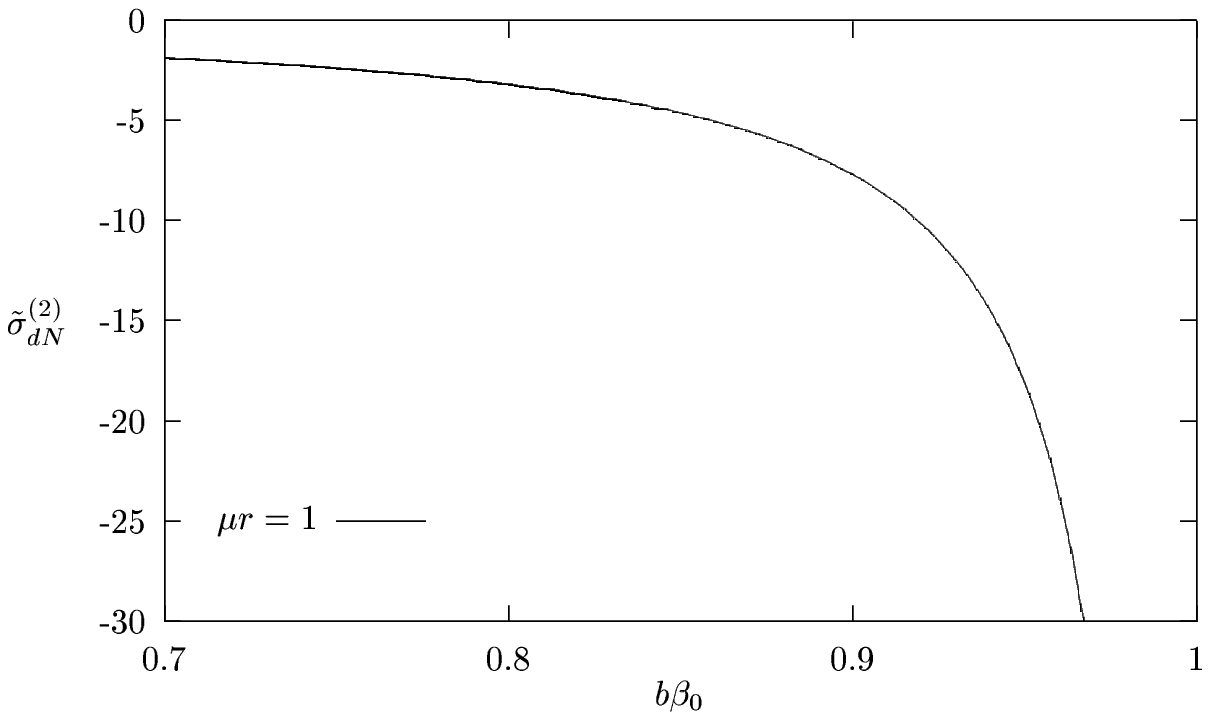}\hfill}}
\caption[]{The Borel structure of $  \tilde{\sigma}_{dN}^{(2)}(r;b)$,
 for  $ \mu r =  1$. The graphs show intermediate $b \beta_0$ and $b
  \beta_0 \rightarrow 1$ behaviour.}
\label{conv_2_b}
\end{figure}

 In section 4, we demonstrated that the exponential nature of
 $\sigma^{(0)}$ at large $\mu r$ was responsible for the Borel
 singularity structure observed in the first convolution. The power
 like behaviour of $  \tilde{\sigma}_{dN}^{(1)}(r;b)$ at large
 $\mu r$ does not imply that new singularities will appear in the
 Borel plane to the left of $b \beta_0 = 1$. It is the action of the
 convolution in $b^\prime$ in each subsequent iteration which
 guarantees this. For large $r^\prime$, the convolution integral
 scales as $r^{\prime(2 b^\prime \beta_0 - 1)}$ from the kernel, $r^{\prime 2
 ((b-b^\prime)\beta_0 -1)}$ from the behaviour of 
 $ \tilde{\sigma}_{dN}^{(1)}(r^\prime ;b)$ and the integration introduces a 
 factor of $r$; the result being that the second convolution also exhibits a
 power dependence of the form given in eq.~(\ref{asyiter1}) for large
 $r$. Indeed this large $r$ behaviour is sufficient to predict the
 position of the leading Borel plane singularity for subsequent
 iterations. Since the infrared singularities are determined by the
 infrared (large $r$) behaviour of the function on which the kernel
 acts, we can consider 
 \begin{equation}
 \int_{0}^{b} \d b^\prime \int \d r^\prime
 \tilde{{\mathcal K}} (r , r^\prime ,b-b^\prime)
 ({r^\prime}^2)^{b^\prime \beta_0 -1} = \frac{N_c}{\pi} \int_{0}^{b}
 \d b^\prime \chi (b^\prime \beta_0 -1,b-b^\prime) \left(
   \frac{Q^2r^2}{\pi}\right)^{b \beta_0} (r^2)^{b^\prime \beta_0 -1} \, .
 \label{scaling}
 \end{equation} 
 Then, from eq.~(\ref{chiform}) we see that $\chi (b^\prime
 \beta_0 -1,b-b^\prime)$ has a leading singularity at $b \beta_0 = 1$,
 which is where the leading singularity will be found for {\it all}
 subsequent iterations.
 This is demonstrated in fig.~(\ref{conv_2_b}). We also note that the
 convolution in $b$ leads to a linear fall off as $b \beta_0
 \rightarrow 0$. For further iterations of the kernel it is the
 intermediate $b \beta_0$ range which will be of most significance.
 Although the leading singularity occurs again at $b \beta_0 = 1$ as
 expected from the discussion above, the shape of the distribution is
 broader. This is because for the second and subsequent iterations the
 leading singularity is converted into a cut with branch point at $b
 \beta_0 = 1$ and receives singular contributions from both the
 virtual correction and real emission part of the kernel.
 
 The generation of each iteration is a numerically intensive task.
 After generating a sufficient number of iterations, we will be in a
 position to solve eq.~(\ref{d3}) and construct the structure
 functions numerically.  As explained above, we do not expect the
 position of the leading singularity to shift.  However, the behaviour
 of the wave function in the vicinity of the singularity contains
 important information about the contribution to structure functions
 from the infrared regions of transverse momentum space. We will
 report on findings for the structure functions in a future
 publication.

 \section{Summary}

 Although there exist numerous parametrisations of experimental data for 
 semihard processes that include non-perturbative effects in plausible ways, 
 here we have attempted to study in a systematic fashion the emergence 
 of non-perturbative corrections as signaled by perturbation theory itself. 
 For small-$x$ D.I.S. we have worked within the dipole cascade formalism and 
 introduced the running coupling in a self-consistent way. 
 To study the resulting loss of scale invariance we considered the dipole 
 evolution equation in Borel ($b$) space. The variable $b$ provides a 
 measure of deviation from the conformal limit $b=0$, where we have shown 
 that the evolution equation reproduces the well known BFKL result.

 We have identified the Borel singularities of the dipole density and have 
 studied how these singularities change once the dipole density is 
 convoluted with the dipole-nucleon cross section. 
 In this case the evolution equation generates leading singularity which 
 is a branch cut at $b \beta_0 = 1$. This was established by numerical 
 calculation for D.I.S. at small $x$. Such a singularity indicates the 
 presence of $1/Q^2$ power corrections to the small-$x$ structure functions 
 coming from the emission of dipoles of large transverse size. 
 These correspond to `ladder gluons' of small transverse momentum 
 in the BFKL formalism. Hence our approach is suitable for studying 
 the region of diffusion of transverse momentum towards the infrared.   
 Moreover, it would be interesting to see how the scale dependence of 
 the interactions in the dipole cascade modify the dependence of the 
 structure functions on $1/x$.

 {\it Acknowledgement:} The authors would like to thank G. Ross for useful
 discussions. One of us (KDA) acknowledges the financial support of
 PPARC.

\newpage

 \end{document}